# Electrochromism in Electrolyte-Free and Solution Processed Bragg Stacks

Liliana Moscardi[1,2,†], Giuseppe M. Paternò[2,†*], Alessandro Chiasera[3], Roberto Sorrentino[1,2], Fabio Marangi[1,2], Ilka Kriegel[4], Guglielmo Lanzani[1,2], Francesco Scotognella[1,2,*].

*1 Department of Physics, Politecnico di Milano, Piazza Leonardo da Vinci 32, 20133 Milano, Italy*
*2 Center for Nano Science and Technology@PoliMi, Istituto Italiano di Tecnologia (IIT), Via Giovanni Pascoli, 70/3, 20133, Milan, Italy*
*3 Istituto di Fotonica e Nanotecnologie IFN – CNR, , Via alla Cascata, 56/C, 3812, Povo – Trent, Italy*
*4 Department of Nanochemistry, Istituto Italiano di Tecnologia (IIT), via Morego, 30, 16163 Genova, Italy*

**ABSTRACT**

Achieving an active manipulation of colours has huge implications in optoelectronics, as colours engineering can be exploited in a number of applications, ranging from display to lightning. In the last decade, the synergy of the highly pure colours of 1D photonic crystals, also known as Bragg stacks, with electro-tunable materials have been proposed as an interesting route to attain such a technologically relevant effect. However, recent works rely on the use of liquid electrolytes, which can pose issues in terms of chemical and environmental stability.

Here, we report on the proof-of-concept of an electrolyte free and solution-processed electrochromic Bragg stack. We integrate an electro-responsive plasmonic metal oxide, namely indium tin oxide, in a 1D photonic crystal structure made of alternating layers of ITO and $TiO_2$ nanoparticles. In such a device we observed 15 nm blue-shift upon application of an external bias (5 V), an effect that we attribute to the increase of ITO charge density arising from the capacitive charging at the metal oxide/dielectric interface and from the current flowing throughout the porous structure. Our data suggest that electrochromism can be attained in all-solid state systems by combining a judicious selection of the constituent materials with device architecture optimisation.

*[†]These authors contributed equally to this work.*

*[*]Address all correspondence to Francesco Scotognella E-mail: francesco.scotognella@polimi.it and Giuseppe M. Paternò Giuseppe.paterno@iit.it*

# INTRODUCTION

The tunability of structural colours in photonic crystals can be exploited to build-up optical active components that can be utilised in a wide range of applications, such as displays,[1,2] colour-changing inks[3,4], smart windows,[5] optical cavities,[6–9] solar cells[10] and sensors,[11–14] among others. In brief, photonic crystals (PhCs) are periodic structures in which there is a repetition of materials with different refractive indices along one, two or three spatial dimensions.[15,16] Such a periodicity generates a forbidden gap for photons,[17] the so-called photonic band gap (PBG). This in turns confers to PhCs structural reflection colours[18,19] without the presence of dyes or pigments, an effect that is widely found in Nature.[20,21]

In this context, one dimensional photonic crystals (1D PhCs, or Bragg Stacks) lend themselves to easy integration with stimuli responsive elements, thanks to the easiness of fabrication of such multi-layered structures.[15,22,23] Responsivity can be achieved through different functionalities, such as liquid crystals,[24] dielectric nanoparticles,[25–28] mesoporous structures[6,10,29–32] and chemical groups,[33,34] which are sensitive to chemical and physical external stimuli.[24,35–38] However, most of the recent works in the field of electrochromic Bragg mirrors (ECBM) rely on the use of liquid electrolytes that can pose issues in terms of chemical and environmental stability.[5,42–44] Among the various degrees of freedom to be exploited to engineer the PhC's optical response, electrochemical doping of plasmonic materials represents a promising approach to modulate the refractive index contrast and, hence, the structural colour without the need to replace the constituent media.[39] For this purpose, the use of heavily doped metal oxide nanoparticles with plasmonic response in the near infrared (NIR) part of the spectrum is of great interest, since the sensibly lower carrier density in these materials compared with classical bulk metals ($10^{21}$ cm$^{-3}$ *vs.* $10^{23}$ cm$^{-3}$) [40,41] allows for a fine control over their dielectric function by means of electro-optical doping. [5,38,42–44]

In this work, we integrate a doped metal oxide widely employed in optoelectronics as transparent electrode namely indium tin oxide (ITO), in a fully solution-processed 1D PhC structure made of alternating layers of ITO and TiO$_2$ nanoparticles. Interestingly, we observe that the tunable optical response of ITO's NIR plasmon resonance via electro-doping , can be translated easily into the visible region thanks to their integration into a photonic structure and the thus obtained modulation of the photonic bandgap. Thus, our results demonstrate that electrochromism in Bragg Stacks can be achieved without in all solid-state architectures without the use of liquid electrolytes.

**EXPERIMENTAL METHODS**

1D PhCs were fabricated by depositing sequentially the constituent $TiO_2$ and ITO nanoparticles via spin casting deposition. $TiO_2$ (Gentech Nanomaterials, average size 5 nm) and ITO (GetNanoMaterials, average size 20 nm) nanoparticles were suspended in Milli-Q® water (10 wt%). The colloidal dispersions were then sonicated for 2 h at 45 °C (Bandelin SONOREX Digital 10 P) in order to improve homogeneity, and filtered with a 0.45 μm PVDF filter. Fluorine tin oxide substrates (FTO, Xop Física S. L.) that were used as bottom electrode, were previously cleaned via sonication bath in isopropanol and acetone for 10 min., and subjected to an oxygen plasma treatment (Colibrì Gambetti) for 10 minutes. 1D PhCs were fabricated on top of such FTO substrates by means of spin-casting deposition of the aqueous dispersions (Laurell WS-400-6NNP-Lite, 2000 rpm), yielding 5 $TiO_2$/ITO bilayers. Finally, we sandwiched the structure with a bare FTO substrate acting as a top electrode using binder clips to hold the structure in place, and then contacted them electrically with copper wires attached to the FTO electrodes by using silver paste. To minimize short circuits in the porous structure, we also fabricated photonic structures integrating two relatively thick (1 μm) layers of $SiO_2$ at the top and bottom of the multilayer. Such dielectric layers were deposited via radio-frequency sputtering (RF), employing the method described previously in reference.[23] For transmission measurements, we used a Perkin Elmer Lambda 1050 WB spectrophotometer, and external bias was applied by using an Agilent 6614C source-meter.

Scanning electron microscopy (Tescan MIRA3) and profilometry (Veeco Dektak 150) were performed to investigate the structure and morphology of the samples.

## RESULTS AND DISCUSSION

In Figure 1a we show the SEM cross-section of the multi-layered porous photonic structure deposited on top of a conductive FTO substrate, consisting of 5 bilayers of $TiO_2$ / ITO spin-cast from their colloidal dispersions. Each layer displays a thickness of 70 nm and 150 nm for $TiO_2$ and ITO respectively, yielding a total thickness of ≈ 1.1 µm. The periodic alternation in 1D of the constituent materials exhibiting different refractive indices (2.6 and 1.9 for $TiO_2$ and ITO in bulk, respectively) gives rises to the photonic band gap and, hence, to the structural reflection colour that in our case lies in the red region (Figure 1b). The photonic structure was then sandwiched with a top bare FTO substrate and clipped with a simple paper binder to ensure mechanical stability during the measurements. Two copper wires were attached to the top and bottom electrodes by using silver paste as reported in Figure c. A simplified sketch of the experimental architecture is reported in Figure 1d, as well as the electro doping effect of the Bragg stack that we aim to achieve in our experiment. An accumulation of charges at the doped semiconductor/$TiO_2$ interface with an applied electric field leads to an effective increase of the charge density contributing to the plasma frequency in ITO (eq. 1):

$$\omega_p = \sqrt{\frac{Ne^2}{m^*\varepsilon_0}} \qquad (1)$$

where $N$ is the charrier density, $e$ is the electron charge, $\varepsilon_0$ is the dielectric constant in the vacuum and $m^*$ in the effective mass. According to the Drude model, [45] the plasma frequency is linked to the complex dielectric function (eq. 2):

$$\varepsilon(\omega) = \varepsilon_1(\omega) + i\varepsilon_2(\omega) \qquad (2)$$

Considering that the dielectric constant is closely related to the refractive index $n$:

$$n = \sqrt{\varepsilon\mu} \qquad (3)$$

with $\mu$ being the magnetic permeability of the medium, which is usually 1 in the visible spectral range, we can therefore conclude that the optical properties of photonic crystals are linked to the frequency dependent dielectric function as described by the Bragg-Snell law:[46]

$$\lambda_{max} = \frac{2}{m} d \sqrt{n_{eff}^2 - sin^2\theta} \qquad (4)$$

where $\lambda_{max}$ is the PBG wavelength, $m$ is the order of the Bragg diffraction, $d$ is the lattice constant, $n_{eff}$ is the effective refractive index between the two materials and $\theta$ is the angle of the incident with the respect to the plane. According to this, one can manipulate the structural colours of PhCs by modifying the charge density of the consistent materials, *i.e.* via photo/electro doping

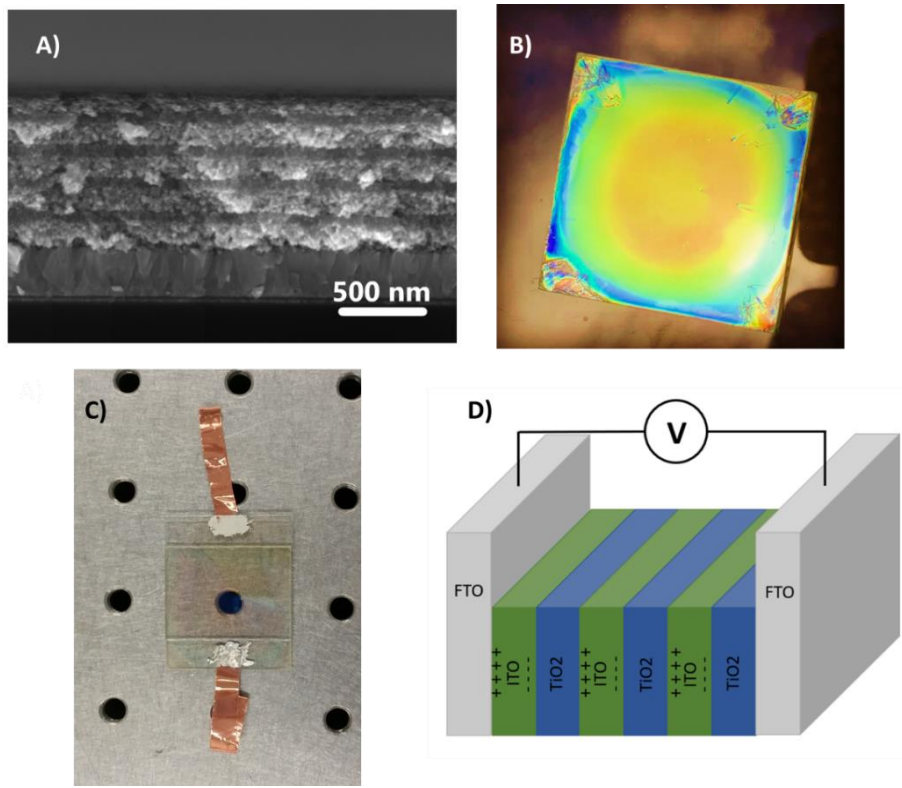

**Figure 1.** (A) SEM image of the multilayered 1D PhC, which consists of alternating layers of $TiO_2$ and ITO nanoparticles (5 bilayers) deposited via spin-casting deposition from their colloidal dispersions on top of an FTO substrate acting as bottom electrode. The average sizes of the nanoparticles dispersion are 5 nm and 30 nm for $TiO_2$ and ITO, respectively. This explains both the higher thickness and roughness of the ITO than the $TiO_2$ layer. (B) Picture of the 1D photonic crystal highlighting the vivid reflection colour of the multilayered structure. (C) Picture of the device architecture. Copper wires were attached to the FTO bottom and top substrate using silver paste. We also made use of clip binders to hold the structure in place and ensure electrical continuity along the vertical direction (now shown here). (D) Depiction of the electro-doping effect that would translate into a modification of the structural colour, as formalized in equations 1-4. Note that the actual devices consist of 5 bilayers, while in the sketch we present 3 bilayers for simplicity.

With this in mind, we proceeded to the electro-doping experiment in which we applied an external bias to the composite structure. The transmission spectrum reported in Figure 2a presents two different contributions, namely: i. a transmittance dip centred at around 600 nm that corresponds to the photonic band gap (PBG); ii. a broad feature in the near infrared that is related to the ITO nanoparticles plasmon resonance.[35] We observed a relatively small but noticeable decrease of the ITO transmittance already at the lowest bias (0.5 V) that does not increase appreciably further upon application of higher voltages. We attribute this to the enhanced carrier density in ITO, as it has been already noticed in photo-doped ITO/SiO$_2$ PhCs.[35] If we focus on the PBG spectral region, we can see that the electro-doping causes a clear blue-shift (15 nm) of the band-gap, with most of the effect (10 nm, 66% of the overall shift) occurring already at 0.5 V. A further increase of the bias led to only minor modifications. Above 5 V short circuits were induced, likely due to the porous structure and to the presence of pinholes in the rough ITO layer causing direct electrical contact between the top and bottom electrode. Morphological investigation over the ITO layer corroborates such hypothesis, as the SEM image highlights the tendency of ITO nanoparticles to assemble into a discontinuous morphology (Figure S2). Furthermore, this suggests that polarization charging at the interface between ITO and TiO$_2$ is not the only process that contributes to the electro-doping. Conversely, a net charge current flowing through the whole structure might represent the predominant effect, thanks to the intrinsically doped metal oxide. Although, these results are interesting *per se* to gain insights into the electro-doping phenomena occurring is such hybrid systems, on the other hand this is undesirable for the effective application of these devices in optoelectronics and photonics, as short circuits damage unavoidably the constituent materials and lead to hysteretic effects.

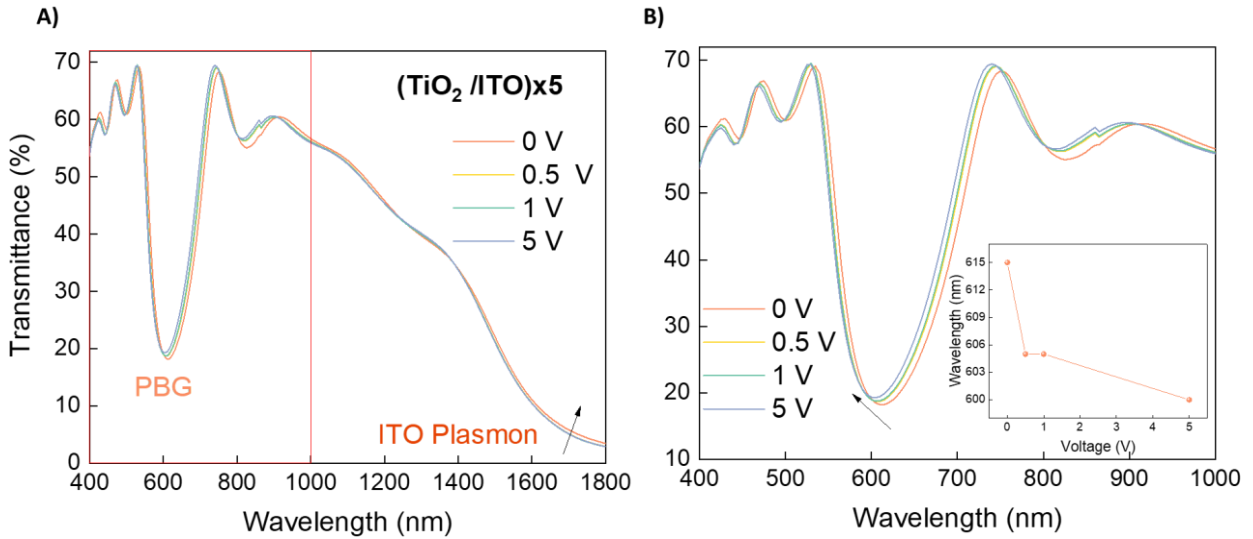

**Figure 1.** (A) Transmission spectrum of the TiO$_2$/ ITO photonic crystal exhibiting a double spectral response: i. the photonic band gap centred at around 600 nm, and the broad plasmon resonance of ITO nanoparticles in the near infrared region. (B) Zoom on the PGB part of the spectrum, showing the spectral shift of the band-gap upon application of the external bias. The inset shows the shift of the central wavelength *vs.* the applied bias. In such a device configuration we could not increase the voltage over 5 V due to the occurrence of short circuiting processes.

For these reasons, to prevent short-circuits and disentangle partially the two effects, we sandwiched the multilayer between two relatively thick (1 μm) dielectric layers of SiO$_2$ deposited via radio-frequency sputtering (see Figure 3a for the SEM cross-section analysis and Figure 3b for a simplified sketch of the device). The insertion of those thick layers reduces necessarily the optical quality of the photonic crystal (Figure 3c), as demonstrated by the appearance of a defect state in the PGB at 500 nm. Nevertheless, such a device architecture permitted to achieve electro-doping and to gain important information on the whole process. Interestingly, we could still appreciate a blue-shift that is lower in magnitude than in the aforementioned architecture (≈ 4.5 nm), and which tends to saturate at 20 V. This in-fact confirms that the capacitive charging is indeed less effective in inducing the overall electro-doping process than the flow of charges due to the electric field itself. Furthermore, in this case we were able to apply a sensibly larger bias (80 V) without any short circuits and, importantly, we noted a partial reversibility of the electro-doping process (40%). Note that we observed full reversibility after two days from the initial experiment, denoting a relatively long discharging time for the dielectric structure (Figure 3d).

To validate such an effect, we also calculated the transmission spectra of the two photonic architectures as a function of ITO charge carrier density (Fig. S1a,b, see supplementary information for the details about the calculation). We indeed observed a blue-shift upon a 60% increase of the ITO charge carrier density, meaning that electro tuning of the overall photonic read-out can be achieved also without the use electrolytes.

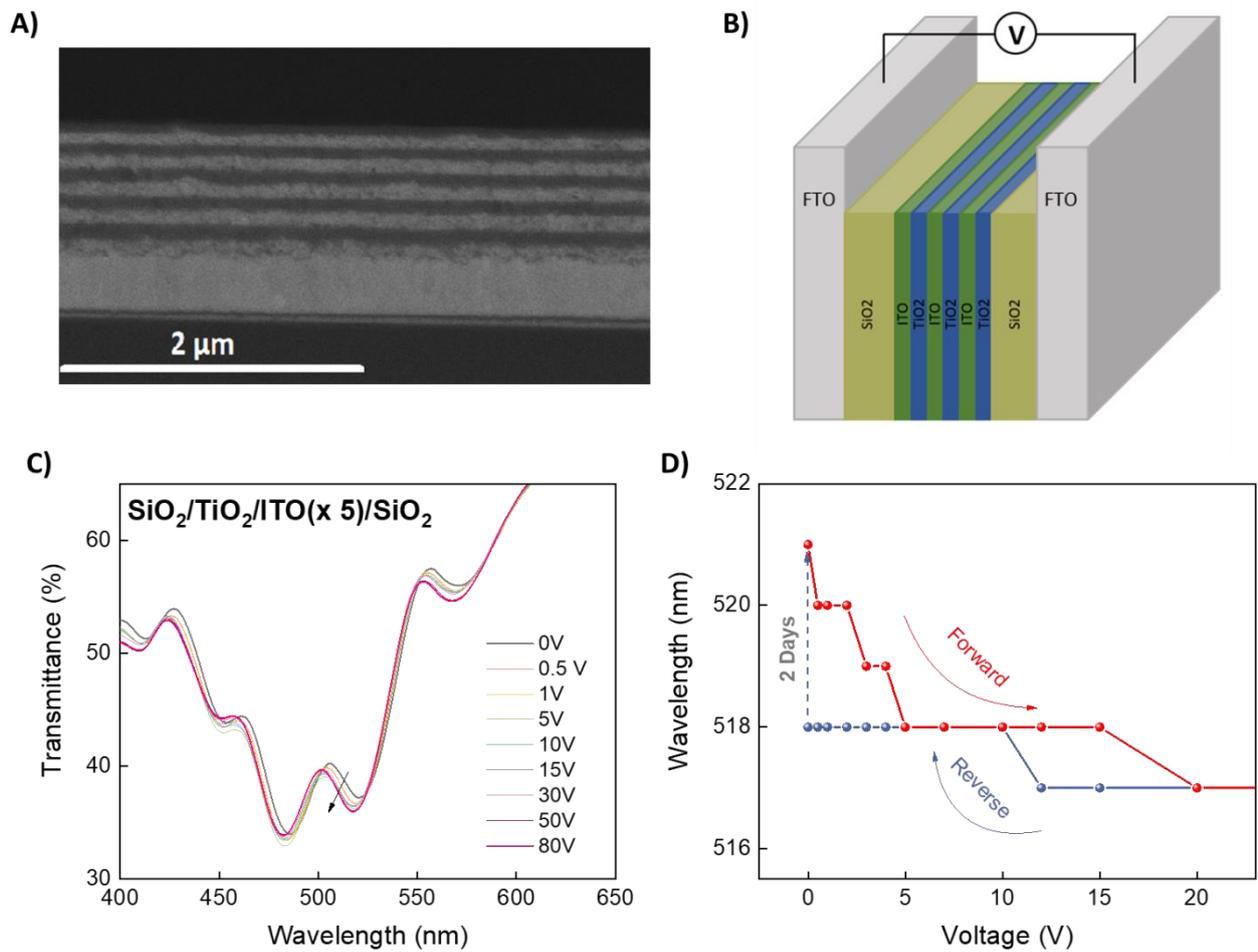

**Figure 2.** (A) SEM cross-section of the photonic multilayer deposited on top of a ≈ 1 μm SiO₂ layer. B) Sketch of the device architecture in which the photonic multilayer is sandwiched between two thick (1 μm) layers of SiO$_2$, to prevent short circuits and minimise the amount of current flowing throughout the structure. Note that the actual devices consist of 5 bilayers, while in the sketch we present 3 bilayers for simplicity. (C) Transmittance spectrum of the structure upon application of an external bias. In this case, the presence of the dielectric materials permitted to increase the external voltage (0 – 80 V). (D) Shift of the central wavelength *vs.* the applied bias both in forward and in reverse mode. Full reversibility of the process was verified after two days from the electro-doping measurement.

## CONCLUSIONS

In this work, we have shown a proof-of-concept of electrochromic Bragg stacks without the use of liquid electrolytes. Our approach, consists in the integration of electro-responsive plasmonic materials in the photonic structure, whose frequency dependent dielectric function can be modulated by application of external stimuli, such as photo/electro doping. This can be translated into a shift in the photonic response, in accordance with the Bragg-Snell law. In particular, we fabricated 1D photonic crystals made of alternating layers of $TiO_2$ and ITO spin-cast from their colloidal dispersions, which display a blue-shift of 15 nm upon application of 5 V. We attribute this to the electro-doping of ITO layers originating from two main contributions: i. the capacitive charging at the ITO/$TiO_2$ interface; ii. the flow of charges due to the electric field itself. However, no reversibility was achieved due to short circuits phenomena likely arising from pinholes in the porous structure. To minimise such latter effect, we sandwiched the structure between two thick dielectric layers of $SiO_2$, observing a relatively lower shift ($\approx$ 4.5 nm) but reversibility of the spectral response. These are promising results for further improvements of the electro-responsivity effects. For instance, we are currently investigating a number of doped metal oxide with tunable plasmon frequency (i.e $WO_3$ among others), while optimising the device structure. These devices can represent novel all-solid state electro-optical platforms that enable active manipulation of colours without the use of liquid electrolytes.

## ACKNOWLEDGEMENTS

F.S. and I.K. thank the European Research Council (ERC) under the European Union's Horizon 2020 research and innovation programme (grant agreement No. [816313] and No. [850875]). GMP and GL acknowledge financial support from Fondazione Cariplo, grant n° 2018-0979 and grant n° 2018-0505.

# Electrochromism in Electrolyte-Free and Solution Processed Bragg Stacks

Liliana Moscardi[1,2][†], Giuseppe M. Paternò[2][†][*], Alessandro Chiasera[3], Roberto Sorrentino[1,2], Fabio Marangi[1,2], Ilka Kriegel[4], Guglielmo Lanzani[1,2], Francesco Scotognella[1,2,*].

*1 Department of Physics, Politecnico di Milano, Piazza Leonardo da Vinci 32, 20133 Milano, Italy*

*2 Center for Nano Science and Technology@PoliMi, Istituto Italiano di Tecnologia (IIT), Via Giovanni Pascoli, 70/3, 20133, Milan, Italy*

*3 Istituto di Fotonica e Nanotecnologie IFN – CNR, , Via alla Cascata, 56/C, 3812, Povo – Trent, Italy*

*4 Department of Nanochemistry, Istituto Italiano di Tecnologia (IIT), via Morego, 30, 16163 Genova, Italy*

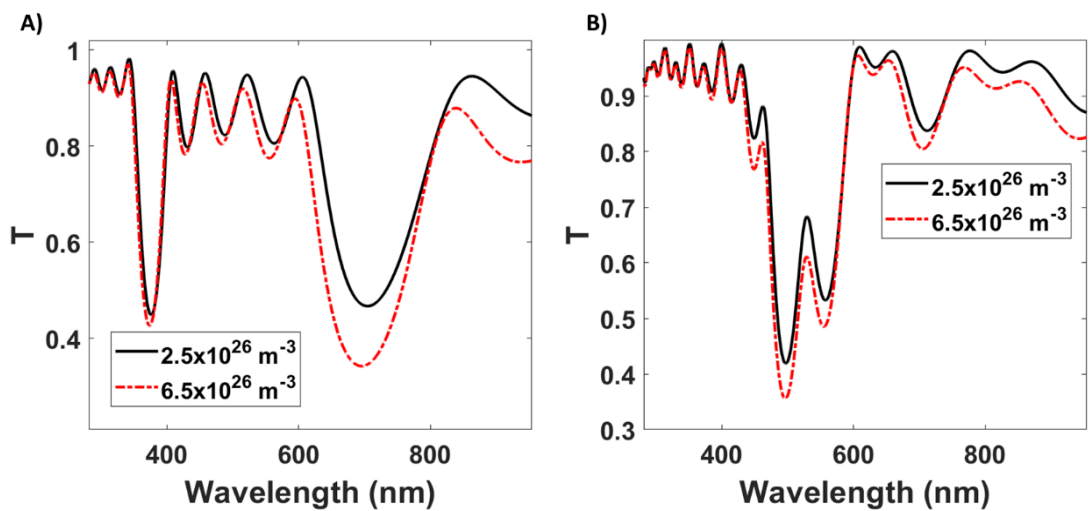

**Figure 3.** Calculated transmission spectrum for the $(TiO_2/ITO)_{x5}$ (a) and for the $SiO_2/(TiO_2/ITO)_{x5}/SiO_2$ (b). We employed the transfer matrix method to model the alternating refractive indexes of the periodic structure combined with the Maxwell− Garnett effective medium approximation for the description of the effective refractive indexes of the TiO₂/ITO. Furthermore, we used the Lorentz−Drude model to account for the ITO plasmonic contribution to the overall dielectric response of the device.

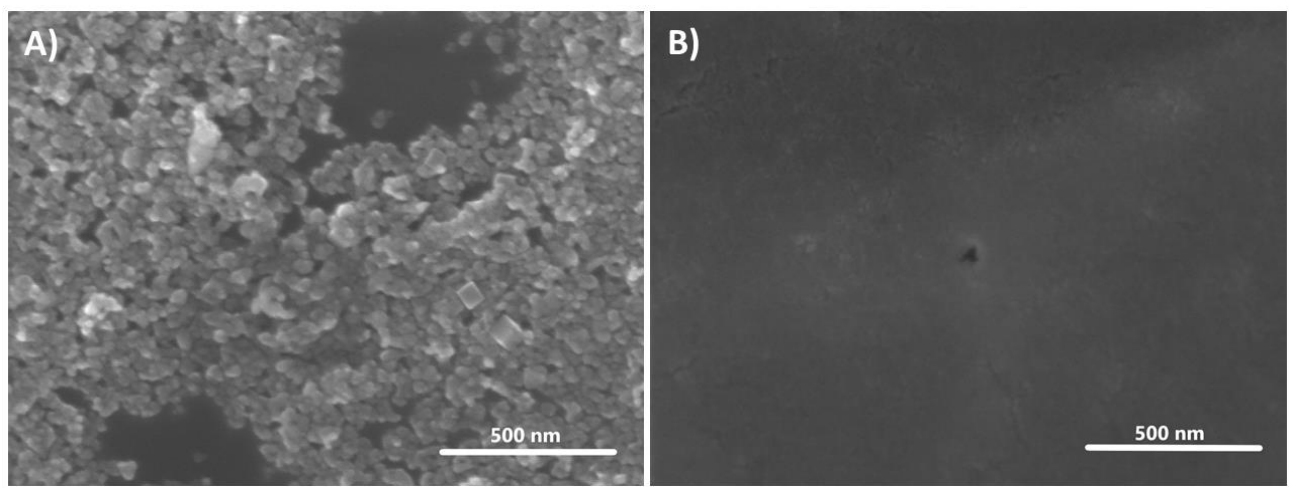

**Figure S2.** SEM images of A) ITO and B) TiO2 layers deposited via spin-casting from the colloidal dispersions.